\documentstyle[12pt]{article}
\begin{document}
\vspace*{3cm}
\def\baselinestretch{2}
\begin{center}{\large \bf The Upper Critical Dimension\\
of the Abelian Sandpile Model}\\  \vspace{10mm}{\bf V.B.Priezzhev}\\
\vspace{5mm}{\it Laboratory of Theoretical Physics, Joint
Institute for Nuclear Research, Dubna, 141980, Russia}
\end{center}
\vspace{10mm}
{\bf Abstract.}
The existing estimation of the upper critical dimension of the Abelian
Sandpile Model is based on a qualitative consideration of avalanches
as self-avoiding branching processes. We find an exact representation
of an avalanche as a sequence of spanning sub-trees of  two-component
spanning trees. Using equivalence between chemical paths on the spanning
tree and loop-erased random walks, we reduce the problem to
determination of the fractal dimension of spanning sub-trees. Then,
the upper critical dimension $d_u=4$ follows from Lawler's theorems for
intersection probabilities of random walks and loop-erased random walks.
\vspace{30mm} \newline
\vspace{10mm} \newline
\newpage
\section{Introduction}
The standard sandpile model introduced in \cite{BTW} became of
mathematical interest after the paper \cite{Dhar} by Dhar who discovered
its Abelian structure. The model is defined on a finite hypercubic
$d$-dimensional lattice. Each site $i$ is characterized by a nonnegative
integer variable $z_i$ called the "height". If $z_i < 2d$ for all $i$,
the sandpile is said to be stable. A vertex is picked at random and
its height is increased by one. If $z_i \geq 2d$, then the site is
unstable and topples giving one particle to each of its neighbors
which in their turn, can be unstable and topple. The process
called "avalanche" continues until all sites become stable and then
a new particle is added to the lattice. If the avalanche reaches the
boundary, sand disappears in a sink connected with all boundary sites.

The "size" of an avalanche may be measured by the total number of
topplings $s$, the number of distinct sites toppled $a$, the diameter
of the region affected by avalanche $r$, and duration of avalanche $t$.
It is generally believed, the probability that the avalanche has size
$x (x=s,a,r,t)$ varies asymptotically
as $x^{-\tau}(\tau=\tau_s,\tau_a,\tau_r,\tau_t)$
when $x \rightarrow \infty$. The mean-field value of the exponent
$\tau_s$ obtained from exact solutions on the Bethe lattice \cite{DM}
and on the full graph \cite{JL} is $3/2$. It is expected that
$\tau_s=3/2$ also for ${\bf Z}^d$ when $d>d_u$ where $d_u$ is the upper
critical dimension.

The first attempt to find $d_u$ was made by Obukhov \cite{O} soon after
the sandpile model was proposed. Briefly, Obukhov's arguments can be
summarized as follows.

(i) Consider changes resulting from an avalanche propagating through the
lattice. If the system is in the recurrent state, the average sensitivity
to a new excitation does not change. During the avalanche, the sites
which have already toppled have heights lower than average, whereas
the heights in the neighboring sites are larger than average. Hence, the
previously activated sites repulse the new activation process. This
situation resembles the True Self-Avoiding Walk (TSAW).

(ii) The essential difference between the sandpile and TSAW
is a possibility of branching the activation process. Thus,
the avalanche can be pictured as a self-avoiding branching process.

Using the renormalization group and $\epsilon$-expansion, Obukhov \cite{O}
calculated one-loop corrections to the mean-field theory and came to the
value $d_u=4$. This conclusion was supported by Dias-Guilera
\cite{DG}  who
analyzed non-linear stochastic differential equations
derived from the models with continuously distributed heights \cite{Zh}.
Later on, Christensen and Olami \cite{CO} suggested $d_u=6$ from an analogy
between spreading of avalanches and  percolation. The mean-field
treatment of a self-organized branching process was discussed in \cite{ZLS}.

The proof of correspondence between the branching TSAW and an avalanche
process in the sandpile model meets two considerable difficulties.
First, a typical avalanche contains multiple topplings which violate
self-avoidance of branches. In higher dimensions, multiple topplings
are suppressed \cite{L},\cite{LU}. However, then, the second problem
arises. The branching process corresponding to an avalanche depends
deterministically on a recurrent configuration where it was initiated.
All recurrent states in the Abelian sandpile model have equal probabilities
\cite{Dhar}.
If self-avoiding branching processes having equal number of steps, have
equal statistical weights, they can be considered as lattice trees also
known as branched polymers. The upper critical dimension of branched
polymers  $d_u=8$ was predicted in \cite{LI} and then
rigorously determined in \cite{BFG},\cite{TH},\cite{HS}.

Thus, avalanches either do not correspond to the branching TSAW due to
multiple topplings, or do not have the specific TSAW weights due to
equal probabilities of recurrent states.

The computation situation is not less controversial. For the most part
of numerical experiments, the accuracy is not sufficient to distinguish
between non-trivial and mean-field values of critical exponents in high
dimensions. Only high-statistics data on large lattices provide some
information. Grassberger and Manna \cite{MG} investigated critical exponents
for $d \leq 5$ and concluded that $d_u=4$. L{\"u}beck and Usadel \cite{LU}
found $d_u=4$, mentioning that the largest considered size of the system
$(L=80)$ for $d=4$ is too small and the corresponding avalanche distribution
exhibits a very narrow power-law interval. In his paper \cite{L}, L{\"u}beck
pointed an important role of logarithmic corrections to the scaling at
$d_u=4$.

At the same time, Chessa et al. \cite{CMVZ} performed a numerical study
of critical exponents in dimensionality ranging from $d=2$ to $d=6$ and
observed the mean-field behavior only in $d=6$ excluding, therefore, $d_u=4$.

Very recently, Vespignani et al \cite{alex} have derived $d_u = 4$ from
a phenomenological field theory, reflecting the symmetries and
conservation laws of sandpiles.

An apparent inconsistency between  the simple self-avoiding branching process
and real avalanches, as well as contradictory numerical results of
different groups, set to find a more transparent proof of the upper
critical dimension for sandpiles.

In this paper, we prove the upper critical dimension $d_u=4$ using
Lawler's theorems \cite{Law} for intersection probabilities of random
walks and loop-erased random walks. We introduce again the self-avoiding
branched polymers for description of avalanches and show that
avalanches are spanning subtrees embedded into a spanning tree of the whole
lattice rather than usual lattice trees. The problem of fractal dimension
of avalanches is reformulated as that for the spanning subtrees.
Using Majumdar's result \cite{M} of the equivalence between the
chemical path on a spanning tree and the loop-erased random walk \cite{Law1},
we reduce the problem to estimations of intersection probability between
random walks and loop-erased random walks.

\section{The Model}

We consider the sandpile model on the $d$-dimensional hypercube
$\Lambda \subset {\bf Z}^d$. Elements of the state space
$\{0,1,2,...(2d-1) \}^{\Lambda}$ are called
stable configurations and are denoted
by $C$. The value $C(i)=z_i, i \in \Lambda$ is the height of the sandpile
at the site $i$. Given a configuration $C$ and a lattice site $i$,
$a_i C$ is the stable configuration obtained by adding a particle at $i$,
and relaxing the system by topplings at all unstable sites $j$, $z_j \geq 2d$.
On toppling at the site $j$,
\begin{equation}  \label{1}
z_{i} \rightarrow z_{i} - \Delta_{ij}\hspace{1cm}i \in \Lambda
\end{equation}
Elements of the matrix $\Delta$ are: $\Delta_{ii}=2d$ for all
$i \in \Lambda$;  $\Delta_{ij}=-1$ for all bonds $(i,j)$, $|i-j|=1$.

The operators $a_i$ commute with each other \cite{Dhar}

\begin{equation}  \label{2}
[a_{i},a_{j}]=0  \hspace{1cm}i,j \in \Lambda
\end{equation}
This allows one to define the identity operator \cite{Dhar}

\begin{equation} \label{3}
I_i = \prod_{j \in \Lambda} a_j^{\Delta_{ij}} \hspace{1cm} i \in \Lambda
\end{equation}
and the equivalence relation between two configurations
$C^{'}$ and $C^{''}$
\begin{equation} \label{4}
C^{'}(j) = C^{''}(j) + \sum_i n_i\Delta_{ij} \hspace{1cm} j \in \Lambda
\end{equation}
where $n_i, i \in \Lambda$ are integers. The equivalence means that
both $C^{'}$ and $C^{''}$ tend to the same stable configuration after
topplings at all unstable sites.

The set of stable configurations $\{ C \}$ splits into two subsets:
$\{ C \} = \{ C \}_R \bigcup \{ C \}_T$   where   $\{ C \}_R$ is the
recurrent set and $\{ C \}_T$ is the transient set of the sandpile process.
In $\{ C \}_R$, one can define an inverse operator $a_i^{-1}, i \in \Lambda$.
Then, the operators $\{ a_i \}$ form a finite Abelian group \cite{Dhar}.

Due to Eq.(\ref{4}), each $C \in \{ C \}_R$ is an element of the
super-lattice in the $|\Lambda|$-dimensional Euclidean space.
The basis-vectors of this lattice are the rows of
the matrix $\Delta$. Therefore, the number
of elements in  $\{ C \}_R$ is \cite{Dhar}
\begin{equation} \label{5}
N_R = Det \Delta
\end{equation}
The recurrent configurations in  $\{ C \}_R$ have equal probabilities
$N_R^{-1}$.

Denote by $\partial \Lambda$ the set of boundary sites of $\Lambda$ and
by $B$ the set of elementary cubes of the $d$-dimensional Euclidian space
centered at $i \in \partial \Lambda$. We choose a site
$\hat{i} \in {\bf Z}^d$ not belonging to $\Lambda$ and call it the root.
The faces of cubes from $B$ which can be connected with $\hat{i}$ without
intersections with another face, are called external faces. We connect each
site $i \in \partial \Lambda$ with $\hat{i}$ by $\nu_i$ bonds where
$\nu_i$ is the number of external faces of the cube $i$. The lattice
$\Lambda$ together with $\hat{i}$ and new bonds forms a graph denoted by
$G$. The site $\hat{i}$ is a sink for particles leaving the lattice
$\Lambda$ during the avalanche.

Using the explicit form of the identity operator Eq.(\ref{3}), one can
construct a new identity operator
\begin{equation} \label{6}
I_{\partial \Lambda} = \prod_{i \in \partial \Lambda}a_i^{\nu_i}
\end{equation}
$I_{\partial \Lambda}$ says that adding $\nu_i$ particles to each
boundary site triggers an avalanche of a special form: each site
$i \in \Lambda$ topples exactly once and the initial configuration $C$
remains unchanged. One can use $I_{\partial \Lambda}$ to construct a
graph representation for any $C \in \{ C \}_R$ by the so-called "burning
algorithm" \cite{MD}. Consider the topplings initiated by
$I_{\partial \Lambda}$ as a fire starting at $\hat{i}$ and burning
sequentially all lattice sites. Once the rules of propagation of fire
are fixed, the set of bonds along which fire propagates forms a spanning
tree $T$ of the graph $G$. There is one-to-one correspondence between
 $ \{ C \}_R$ and the set of spanning trees $T$. Then, Dhar's formula,
Eq.(\ref{5}), for $N_R$ coincides with Kirhhoff's theorem for the number of
spanning trees \cite{Harary}.

The matrix $\Delta$ in Eq.(\ref{5}) is the minor of the Laplacian matrix
$\Delta_G$ of the graph $G$ corresponding to the element
$(\Delta_G)_{\hat{i} \hat{i}}$. According to Dhar \cite{Dhar}, the Green
function
\begin{equation} \label{7}
G_{ij} = [\Delta^{-1}]_{ij}
\end{equation}
is the average number of topplings at the site $j$ due to a single particle
added at $i$ in a configuration $C \in \{ C \}_R$. For the random walk
defined by the matrix $\Delta$, $G_{ij}$ is the expected number of times
the walk started at $i$ visits the site $j$ until it is trapped by the
absorbing site $\hat{i}$ \cite{Spitzer}.

To find the spanning tree representation for $G_{ij}$, we delete from
$\Lambda$ the elementary cube centered at $i$ and consider its $2d$
faces as a part of the new external boundary. The sites adjacent to $i$
form a new boundary set $\partial^i \Lambda$. Repeating the construction
of a spanning tree by the identity operator, Eq.(\ref{6}), with the boundary
set $\partial \Lambda \cup \partial^i \Lambda $, we obtain a
two-component spanning tree having the roots $\hat{i}$ and $i$ for two
different subtrees.

{\it Proposition 1} proved in \cite{IKP} (see also \cite{IKPP})
reads: For any connected graph $G$ with a fixed vertex $\hat{i}$
\begin{equation} \label{8}
G_{ij} = N^{(i,j)}/|T|
\end{equation}
where $N^{(i,j)}$ is the number of two-component spanning trees having the
roots $\hat{i}$ and $i$, such that both vertices $i$ and $j$ belong to
the same component; $|T|$ is the number of spanning trees on $G$.

\section{Avalanches and Waves}

An avalanche starting at $i$ is the process of transformation
$C \rightarrow a_i C$. Generally, an avalanche consists of multiple
topplings at different sites and has a complicated structure. However,
one can try to decompose it into simpler subprocesses. Suppose
$z_i = 2d$ after adding one particle to $i$. Since the topplings can
be performed in any order, we topple once at $i$, and then topple all
other unstable sites keeping the site $i$ out of the next toppling.
This sub-avalanche is the first wave of topplings. If $i$ is still
unstable, we topple $i$ once again giving rise to the second wave of
topplings. This process is continued until $i$ becomes stable. Thus,
an avalanche is broken into a sequence of waves of topplings.

Three important properties of waves make them useful for the
analysis of avalanche statistics \cite{IKP}. Let $S_k, k \geq 1$ be
a set of sites toppled during the $k$-th wave. Then,

(i) each $j \in S_k$ topples once and only once;

(ii) both $S_k$ and $\Lambda \setminus S_k$ are connected sets;

(iii) $k$-th wave is the last wave of an avalanche iff the initial
site $i \in S_k$ has a neighbor in $\Lambda \setminus S_k$.

The number of waves in the avalanche $C \rightarrow a_i C$ will be denoted
by $n_i(C)$ or simply $n_i$.

The spanning tree representation of waves can be constructed as follows.
Given a configuration $C \in \{ C \}_R$, $P_i(C)$ represents the projection
of $C$ on the lattice  $\Lambda \setminus i$
where the site $i$ is now considered
as the second root. Let $C_k, k \leq n_i$ be an (unstable) configuration
obtained after the $k$-th wave starting at $i$ is completed. Then, $P_i(C_k)$
can be produced from $P_i(C)$ by the operator
\begin{equation} \label{9}
W_k = (\prod_{j\in\partial^i\Lambda}a_j)^k
\end{equation}
acting on the lattice $\Lambda \setminus i$.

Consider the configuration
\begin{equation} \label{10}
P_i(C_{k-1}) = W_{k-1}P_i(C)
\end{equation}
preceeding the $k$-th wave. If one applies to $P_i(C_{k-1})$ the identity
operator
\begin{equation} \label{11}
I_{\partial \Lambda \cup \partial^i \Lambda } =
(\prod_{j\in\partial^i\Lambda}a_j)
(\prod_{j\in\partial\Lambda}a_j^{\nu_j}),
\end{equation}
one obtains a two-component spanning tree representing  $P_i(C_{k-1})$.
On the other hand, the expression in the first brackets is exactly the
operator providing the $k$-th wave. Therefore, the subtree rooted at $i$
and embedded into the subtree rooted at $\hat{i}$, is the graph
representation of the $k$-th wave.

This result gives another proof of Eq.(\ref{8}).
As each wave beginning at $i$ and
involving $j$ corresponds to one toppling at $j$, $N^{(i,j)}/|T|$ is
the average number of topplings at the site $j$ for a single particle added
at $i$. Then, Eq.(\ref{8}) follows from Eq.(\ref{7}).

Due to the graph representation, we come again to a lattice-tree or
a branched-polymer picture of avalanches. However, the essential
difference  with the qualitative Obuhkov arguments is, firstly, that
we associate the lattice trees with waves
but not with the whole avalanches, and
secondly, the lattice trees are conditioned  by the spanning tree
construction.

Let $s_k$ denote the number of sites in the $k$-th wave $S_k$.
The total number of topplings $s$ in an avalanche is
\begin{equation} \label{12}
s = \sum_{k=1}^{n_i}s_k
\end{equation}
where $n$ is the number of topplings at the initial point $i$
or, equivalently, the number of waves in the avalanche. It is expected
that, at least for sufficiently large dimensions $d$, the probability
distribution of $s$ has the scaling form
\begin{equation} \label{13}
P(s;L) \sim s^{-\tau} f(s/L^D)
\end{equation}
where $D$ is the capacity fractal dimension of avalanches. It follows
from Eq.(\ref{12}) that $D$ depends on the fractal dimension of waves
and the number of waves in an avalanche. In $d=2$, the dimension of waves is
$d_w = 2$ and the number of waves is expected to obey the scaling law
\cite{MD}
\begin{equation} \label{14}
\langle n_i \rangle \sim r^y
\end{equation}
Then the exponents $\tau_s,\tau_a,\tau_r,\tau_t$ can be determined in terms
of $y$. In higher dimensions, multiple toppling events occurs more rarely
showing a tendency to decrease $y$. In Section {\bf 7},
we will show that
for $d=4$, $\langle n_i \rangle$ grows with $r$ not faster than
logarithmically. If so, the problem of critical dimension for avalanches
can be formulated as that for waves.

\section{Statistics of Waves}

We start this section with the definition
of the fractal dimension of waves. Consider
the graph representation of a wave as a two-component spanning tree on
the graph $G$ with the roots $i \in \Lambda$ and $\hat{i}$. The wave $S$
is the set of $s = |S|$ sites belonging to the component rooted at $i$.
The radius of the wave is
\begin{equation} \label{15}
R(S) = sup\{ d(i,j):j \in S\}
\end{equation}
where $d(i,j)$ is the distance between the sites $i$ and $j$. We will say
that waves have a fractal growth if
\begin{equation} \label{16}
\langle s \rangle \sim R^{d_{fg}}
\end{equation}
for large $R$, where $d_{fg}$ is the dimension of fractal growth and
the average is taken over all waves of the radius $R$. The fracral growth,
however, does not guarantee the fractal structure by itself.

Surround the point $i$ by a ball of radius $r$ and consider
all sites of $S$ inside the
ball. Denote the number of internal points by $m(r)$. We say that waves
have the fractal density if, given $\epsilon > 0$, there exist
$r(\epsilon)$, $R(\epsilon)$ and a constant $\bar{c} < 1$ such that
\begin{equation} \label{17}
|log\langle m(r) \rangle -d_{fd} log r| \leq \epsilon
\end{equation}
on the interval
$r(\epsilon) \leq r \leq \bar{c}R$ for all $R > R(\epsilon)$.

If $d_{fg} = d_{fd} =d_f$, the fractal is uniform and $d_f$ is the
fractal dimension of waves on the $d$-dimensional lattice. We do not prove
here uniformity of the fractal structure of waves. Assuming it, we will
find the upper and lower bounds for $d_f$.

In the spherically symmetrical case, we can introduce the density of
waves which varies as
\begin{equation} \label{18}
\rho(r) \sim \frac{1}{r^{d-d_f}}
\end{equation}
in the scaling interval $r(\epsilon) \leq r \leq \bar{c}R$.

If $d_f$ is known, one can determine the probability distribution
of waves $P(x), x=s,a,R,t $ ($s=a$ for waves). Indeed, the expected
number of topplings at a point $j$ in an avalanche starting at $i$,
$|i-j|=r$ varies as
\begin{equation} \label{19}
\langle n(r) \rangle \sim \rho(r) \int_{r/\bar{c}}^{\infty}P(x)dx
\end{equation}
where the integral gives all waves having reached the radius $r/\bar{c}$.
On the other hand, $\langle n(r) \rangle$ coincides with the Green
function, Eq.(\ref{7}). The asymptotics of $G(r)$ in the $d$-dimensional
case is
\begin{equation} \label{20}
G(r) \sim \frac{1}{r^{d-2}}
\end{equation}
Comparing with Eq.(\ref{19}),we find that
\begin{equation} \label{21}
P(R) \sim \frac{1}{R^{d_f-1}}
\end{equation}
The probability distribution for $s$ follows from Eq.(\ref{16})
provided that $s \sim R^{d_f}$ when $R \rightarrow \infty$
\begin{equation} \label{22}
P(s) \sim \frac{1}{s^{\tau_s}}
\end{equation}
with $\tau_s = 2-2/d_f$.
The exponent $\tau_s$ reaches its mean-field value $3/2$ when $d_f = 4$.

To proceed with the determination of $d_f$, we need a more elaborate
definition of $\rho(r)$. Consider a wave $S$ with the initial site $i$.
The set of sites not belonging to $S$ is denoted by $\hat{S}$,
$S \cup \hat{S} = \Lambda$. According to Eq.(\ref{15}),
a wave has the radius not less than $R$ if there exists a site
$i^{'} \in S$, $|i-i^{'}| = R$. The wave $S$ has the density less than 1
at the radius $r \leq R$ if there exists a site $\hat{i}^{'} \in \hat{S}$,
$|i-\hat{i}^{'}| = r$. Consider the two-component spanning tree
with the components
corresponding to $S$ and $\hat{S}$. The path $\Gamma(i,j)$ on a tree between
the points $i$ and $j$ is the sequence of bonds
$(i,i_1),(i_1,i_2),...(i_n,j)$.
Any two points belonging to one component can be connected by a unique
path. Then, we can define the density of waves at the radius $r \leq R$
by the conditional probability that, given a path $\Gamma(i,i^{'})$,
$|i-i^{'}| = R$, there exists a path $\Gamma(\hat{i},\hat{i}^{'})$,
$|i-\hat{i}^{'}| = r$,
\begin{equation} \label{23}
\rho(r) = 1 - Prob(\Gamma(\hat{i},\hat{i}^{'})|\Gamma(i,i^{'}))
\end{equation}
The expression in the right-hand side of Eq.(\ref{23})
is the non-intersection
probability of two self-avoiding paths on a tree, called often
"chemical paths". The first path $\Gamma(i,i^{'})$ connects the site
$i$ with the site $i^{'}$ belonging to the wave $S$; the second path
$\Gamma(\hat{i},\hat{i}^{'})$ connects the site $\hat{i}^{'}$ not
belonging to $S$ with the sink $\hat{i}$.
Having no rigorous results for this
probability, we can, however, reduce it to non-intersection
probability between one chemical path and the simple random walk.

For this purpose, we need a generalization of  Proposition 1 to
the
multicomponent case. Consider a connected graph $G$ and select
a subset of its vertices $\{v\}=i_1,i_2,...,i_{\nu} $.
Let $\Delta_{\{v\}}$ be the matrix obtained from the Laplacian
matrix $\Delta_G$ by deleting all rows and columns corresponding
to $i_1,i_2,...,i_{\nu} $. The Green function
\begin{equation} \label{24}
G_{ij}=[\Delta_{\{v\}}^{-1}]_{ij}
\end{equation}
is the expected number of visits of the site $j$ by the random walk
starting at $i$ before it is absorbed at one of the sites
$i_1,i_2,...,i_{\nu} $.

{\it Proposition 2}\hspace{0.5cm}
For any connected graph $G$ with a fixed subset of
vertices $\{v\}$
\begin{equation} \label{25}
G_{ij}=\frac{ N_{\{v\}}^{(ij)}}{|T_{\{v\}}|}
\end{equation}
where $ N_{\{v\}}^{(ij)}$ is the number of $(\nu+1)$-component spanning
trees having the roots $i_1,i_2,...,i_{\nu}$ and $i$ such that vertices
$i$ and $j$ belong to the same component; $|T_{\{v\}}|$ is the number
of $\nu$-component spanning trees on $G$ having the roots
$i_1,i_2,...,i_{\nu} $.

{\it Remark}  If a site $i_{\mu} \in \{v\}$ is an isolated site of
the spanning tree, we consider it as a component consisting of the
single site.

{\it Proof}  Consider the  graph $G^{'}$ which is the graph $G$ with
all sites $i_1,i_2,...,i_{\nu} $ contracted to the single site $i_0$.
Proposition 1 is applicable to $G^{'}$. It relates the Green function,
Eq.(\ref{24}), with the number of two-component spanning trees having
the roots $i_0$ and $i$, such that the sites $i$ and $j$ belong to
the same component. Returning to the graph $G$, we get from each
configuration on $G^{'}$, a spanning tree where the component containing
$i_0$ splits into $\nu$ components. This construction implies
Proposition 2.

In order to treat the right-hand side of Eq.(\ref{23}) as a random walk
probability, we consider the graph $G$ with the subset of vertices
$\{v\}$ coinciding with the set of all sites of the chemical path
$\Gamma(ii^{'})$. Then, by Proposition 2, we have
\begin{equation} \label{26}
G_{\hat{i}^{'}\hat{i}}=\frac{ N_{\{v\}}^{(\hat{i}\hat{i}^{'})}}{|T_{\{v\}}|}
\end{equation}
where  $G_{\hat{i}^{'}\hat{i}}$ is the expected number of visits of the site
$\hat{i}$ by the random walk starting at $\hat{i}^{'}$ and escaping
the chemical path $\Gamma(ii^{'})$.
If $\nu$ is the number of sites in $\Gamma(ii^{'})$, then
$N_{\{v\}}^{(\hat{i}\hat{i}^{'})}$ is the number of $(\nu+1)$ component
spanning trees having sites $\hat{i}$ and $\hat{i}^{'}$ in one component.
The remaining $\nu$ components can be joint into a single component if one
adds to $\nu$ spanning subtrees rooted at the sites of $\Gamma(ii^{'})$,
the path $\Gamma(ii^{'})$ itself. Then,
$N_{\{v\}}^{(\hat{i}\hat{i}^{'})}$ is the number of two-component
spanning trees which have the path $\Gamma(ii^{'})$  in one component
corresponding to the wave $S$ and the path  $\Gamma(\hat{i}\hat{i}^{'})$
in the second component $\hat{S}$ having the root $\hat{i}^{'}$ or,
equivalently, $\hat{i}$.
$|T_{\{v\}}|$ is the number of all one-component spanning trees
containing the path $\Gamma(ii^{'})$.

Along with $|T_{\{v\}}|$, we introduce $N_{\{v\}}^{(\hat{i})}$, the
number of two-component spanning trees containing the path $\Gamma(ii^{'})$
in one component and the root $\hat{i}$ in the second one. By definition,
\begin{equation} \label{27}
Prob(\Gamma(\hat{i}\hat{i}^{'})|\Gamma(ii^{'}))=
\frac{ N_{\{v\}}^{(\hat{i}\hat{i}^{'})}}{N_{\{v\}}^{(\hat{i})} }=
\frac{N_{\{v\}}^{(\hat{i}\hat{i}^{'})}}{|T_{\{v\}}|}
\frac{|T_{\{v\}}|}{N_{\{v\}}^{(\hat{i})}}
\end{equation}
Using Eq.(\ref{26}), we get
\begin{equation} \label{28}
Prob(\Gamma(\hat{i}\hat{i}^{'})|\Gamma(ii^{'}))=
\frac{G_{\hat{i}^{'}\hat{i}}}{G_{\hat{i}\hat{i}}}
\end{equation}
where the Green function $G_{\hat{i}\hat{i}}$ is the expected number
of returns to the initial point for random walks starting at $\hat{i}$
and escaping the path  $\Gamma(ii^{'})$.

Finally, we note that the ratio of Green functions in Eq.(\ref{28})
is the probability
that the random walk starting at $\hat{i}^{'}$
escapes the path $\Gamma(ii^{'})$ and reaches the site $\hat{i}$. This
fact follows from the property of Green functions known for infinite
lattices as the "weak ergodic theorem" \cite{Spitzer}.
Denoting by $F(\hat{i}^{'}|ii^{'})$ the escaping
probability averaged over all paths $\Gamma(i,i^{'})$, we can summarize
the obtained results as

{\it Proposition 3}\hspace{0.5cm} The waves starting
at a site $i$ of the lattice
$\Lambda$ with the sink $\hat{i}$ have the density
\begin{equation} \label{29}
\rho(r) = 1 - F(\hat{i}^{'}|ii^{'})
\end{equation}
at the point $\hat{i}^{'}$, $|i-\hat{i}^{'}|=r$, if the radius of the
wave exceeds $R = |i-i^{'}| \geq r$.

\section{Loop erased random walks}

The loop erased random walk (LERW) has been introduced by Lawler \cite{Law1}
to modify the uniform measure of the usual self-avoiding random walk where
every possible walk of a fixed length is given the same statistical weight.
Let $[x_{0},...,x_{m}]$ be the simple random walk in ${\bf Z}^{d}$. Then
the LERW can be constructed as follows: let $j$ be the smallest value such
that $x_{j}=x_{i}$ for $0\leq i < j \leq m$. Deleting all steps between
$i$ and $j$, we obtain a new walk $[x_{0},...,x_{i},x_{i+1},...,x_{m}]$.
If the new walk is self-avoiding we stop the process; otherwise, we
continue the loop-erasing operation until we get a self-avoiding path.

The distribution for each step of the LERW depends on the entire past
history being equivalent to that given by the transition probability
of the Laplacian random walk \cite{LE},\cite{Law2}. Another equivalence
has been found by Majumdar \cite{M} who considered a model of growing
trees introduced by Broder \cite{Brod}.
The chemical path on these trees corresponds to the LERW. Like the burning
algorithm, Broder's algorithm in the limit of lattice filling
gives spanning trees with equal probability. Therefore, the statistical
properties of the LERW are identical to those of the chemical path on
spanning trees.

Let $R_n(i)$ and $L_n(i)$ be simple and loop-erased random walks starting
at the site $i$ and let $\Pi_i$ and $\Gamma_i$ denote paths of the
walks
\begin{equation} \label{30}
\Pi_i(a)=\{R_n(i): 0\leq n \leq a\}
\end{equation}
\begin{equation} \label{31}
\Gamma_i(a)=\{L_n(i): 0\leq n \leq a\}
\end{equation}
Denote by $\sigma(n)$ the number of steps of the simple random walk needed
to produce $n$ steps of the LERW
\begin{equation} \label{32}
\sigma(n) = sup\{j:L_n(i)=R_j(i)\}
\end{equation}
For any random walk, $\sigma(n) \geq n$. For a fixed $\sigma(n)$,
$\Pi_i(\sigma(n))$ is the path of a finite random walk.
Then, $\Gamma_i(n)$ is the finite
LERW obtained from $\Pi_i(\sigma(n))$ by the loop-erasing procedure.

The escaping probability of a set $A \in {\bf Z}^d$  is defined as
\begin{equation} \label{33}
Es(i,A) = Prob \{R_n(i) \notin A,n=1,2,3,... \}
\end{equation}
where $i$ is the starting point of the random walk.
Putting $A = \Gamma_i(n)$, we can define $a_n$ as
\begin{equation} \label{34}
a_n = E[Es(i,\Gamma_i(n))]
\end{equation}
i.e. $a_n$ is the probability that an infinite independent simple random
walk starting in $i \in {\bf Z}^d$ does not intersect the finite LERW
derived by erasing loops from $\sigma(n)$ steps of an independent random
walk also starting in $i$.

For $d=4$, Lawler \cite{Law53} has found the lower and upper bounds for $a_n$
\begin{equation} \label{35}
\lim_{n\rightarrow \infty} inf\frac{\log a_n}{\log\log n}\geq -\frac{1}{2}
\end{equation}
and
\begin{equation} \label{36}
\lim_{n\rightarrow \infty} sup\frac{\log a_n}{\log\log n}\leq -\frac{1}{3}
\end{equation}
The upper bound Eq.(\ref{36}) was conjectured \cite{Law53} to coincide with
the exact asymptotics. This result would imply that the mean square
distance for LERW's behaves as
\begin{equation} \label{37}
\langle r_n^2\rangle \sim n (\log n)^{1/3}
\end{equation}
whereas $\langle r_n^2\rangle \sim n$  for $d \geq 5$.

It is convenient to bring here the known intersection probability for two
simple random walks \cite{Law}.
Let $R_n(i)$ and $R_n(j)$ be independent random walks
starting at the points $i$ and $j$, $|i-j|=r$. Then, there exist constants
$c_1 > 0$ and $c_2 > 0$ such that

\begin{equation} \label{38}
c_1 \leq Prob\{\Pi_i(n) \cap \Pi_j(n)\neq 0\} \leq c_2
\end{equation}
for $d<4$
\begin{equation} \label{39}
c_1(\log n)^{-1} \leq Prob\{\Pi_i(n) \cap \Pi_j(n)\neq 0\}
\leq c_2(\log n)^{-1}
\end{equation}
for $d=4$
\begin{equation} \label{40}
c_1 n^{(4-d)/2} \leq Prob\{\Pi_i(n) \cap \Pi_j(n)\neq 0\}
\leq c_2 n^{(4-d)/2}
\end{equation}
for $d>4$, if $a\sqrt n \leq |r| \leq b\sqrt n $,
where $a$ and $b$ are positive constants.

The non-intersection probability of a finite random walk and an infinite
random walk starting at the same point $i$ for $d=4$ is \cite{Law},\cite{Dup}
\begin{equation} \label{41}
Prob\{\Pi_i(n) \cap \Pi_i(\infty) = 0\}
\sim c(\log n)^{-1/2}
\end{equation}
Also, for further estimations, we need the intersection probabilities
between a finite random walk starting at $i$ and an infinite random walk
starting at $j$ for $d=4$ \cite{Law86}
\begin{equation} \label{42}
Prob\{\Pi_i(n) \cap \Pi_j(\infty) \neq 0\} \leq c_1
\log (1+1/\alpha)/\log n
\end{equation}
if $\alpha = |i-j|^2/n$.
\section{Upper and lower bounds for density of waves}

Every random walk on a graph $G$ with an absorbing set $\{v\}$ and the sink
$\hat{i}$ is trapped either by $\{v\}$ or by $\hat{i}$. Then, Proposition 3
means that $\rho(r)$ is the intersection probability between the random
walk starting at the site $\hat{i}^{'}, |i-\hat{i}^{'}|=r$ and the chemical
path $\Gamma(i,i^{'}), |i-i^{'}|=R$.

Majumdar \cite{M} has proved that chemical paths correspond to LERW's which,
in turn, can be obtained from simple random walks by the loop-erasing
procedure. Let $R_n(i)$ be the random walk corresponding to the LERW
$L_n(i)$. As the number of steps $\sigma(n)$ in the path
$\Pi_i(\sigma(n))$, Eq.(\ref{30}), needed to reach the radius $R$,
exceeds the number of steps $n$ in the LERW path $\Gamma_i(n)$,
Eq.(\ref{31}), we can estimate $\rho(r)$ as
\begin{equation} \label{43}
\rho(r) \leq Prob\{\Pi_i(\sigma(n)) \cap \Pi_{\hat{i}^{'}}(\infty)\neq 0\}
\end{equation}
In $d=4$, we get from Eq.(42)
\begin{equation} \label{44}
\rho(r) \leq c \log(1+\frac{1}{\alpha})/\log r
\end{equation}
where $\alpha = r^2/\sigma(n) \sim r^2/R^2$. The fractal density of waves
in $d=4$ decays with $r$ at least logarithmically when $\alpha$ is fixed.
In $d \geq 4$, we can use Eq.(\ref{40}) assuming that the asymptotics does
not change when one of the random walks is extended to infinity. Then, we
have
\begin{equation} \label{45}
\rho(r) \leq c r^{4-d}
\end{equation}
From the definition, Eq.(\ref{18}), we can see that the fractal
dimension of waves $d_f \leq 4$ for all $d \geq 4$.

In $d \leq 4$, the intersection probability, Eq.(\ref{38}), and, therefore
$\rho(r)$, are restricted from above by a constant.

To get a lower bound for $\rho(r)$ in $d=4$, we will use the upper bound
Eq.(\ref{36}) for escaping probability $a_n$.
Consider the random walk $R_n(i)$ and decompose its path
$\Pi_i(\infty)$ into two parts:
$\Pi_i(\infty) = \Pi_i(k) \cup \Pi_{\hat{i}^{'}}(\infty)$
where $k$ is the moment of the first hitting into the site $\hat{i}^{'}$
separated from $i$ by the distance $r$: $R_i(k) = \hat{i}^{'},
R_i(l) \neq \hat{i}^{'}, l=1,2,...,k-1 $.
Then, escaping probability $a_n$ for the LERW of length $n$ by the random
walk $R_n(i)$ is the product of escaping probabilities $a_n(i,\hat{i}^{'})$
and $a_n(\hat{i}^{'},\infty)$ before and after the first hitting into the
site $\hat{i}^{'}$
\begin{equation} \label{46}
a_n = a_n(i,\hat{i}^{'})a_n(\hat{i}^{'},\infty)
\end{equation}
Granting that $a_n(\hat{i}^{'},\infty) =
F(\hat{i}^{'}|i i^{'})$, we can write Eq.(\ref{29}) as
\begin{equation} \label{47}
\rho(r) = 1 - \frac{a_n}{a_n(i,\hat{i}^{'})}
\end{equation}
where $n$ is the length of the LERW coinciding with the path
$\Gamma(i,i^{'})$.

The probability $a_n(i,\hat{i}^{'})$ to reach the point $\hat{i}^{'}$
from $i$ for $m$ steps, $m \sim |i-\hat{i}^{'}|^2=r^2$, escaping the LERW
of length $n$ is not less than the probability to avoid the random walk
of length $\sigma(n)$ from which the LERW was obtained by the loop-
erasing procedure, and therefore not less than the probability
$Prob\{ \Pi_i(\infty)\cap\Pi_i(m)=0\}$ to avoid an infinite random
walk. Using Eqs.(\ref{36}) and (\ref{41}) we get from Eq.(\ref{47})
\begin{equation} \label{48}
\rho(r) \geq 1 - c\frac{(\log m)^{1/2}}{(\log n)^{1/3}}
\end{equation}
For any $r \sim m^{1/2}$, we see from Eq.(\ref{37}) that the number
of steps $n$ in the LERW increases with the radius of wave as $R^2$
(with the logarithmic correction) and $\rho(r)$ approaches 1 when
$R \rightarrow \infty $. The only fractal dimension which is consistent
with both upper and lower bounds Eq.(\ref{44}) and Eq.(\ref{48}) is
4 although the fractal density  Eq.(\ref{18}) as well as the probability
distribution of waves Eqs.(\ref{21}),(\ref{22}) need logarithmic corrections.

The lower bound Eq.(\ref{48}) becomes stronger in $d < 4$ because of
decreasing escaping probability $F(\hat{i}^{'}|ii^{'})$. Assuming that
$d_f$ is a non-decreasing function of $d$, we conclude that the fractal
dimension of waves is

\begin{equation} \label{49}
d_f= \left\{
\begin{minipage}{5cm}
$d$ \hspace{1cm} for $d \leq 4$\\
4 \hspace{1cm} for $d > 4$
\end{minipage}
\right.
\end{equation}
According to Eq.(\ref{22}), the upper critical dimension for
waves is $d_u = 4$.

\section{Number of Waves in Avalanches}

In order to derive the upper critical dimension for avalanches
from that for waves, we need some information about the expected number
of waves in an avalanche. Due to the property (iii) of waves mentioned
in Section {\bf 3}, the probability that a wave starting at the site $i$
is the last wave in an avalanche is proportional to the density of "holes"
in the wave at its origin, i.e. the probability that a site in the vicinity
of $i$ does not belong to the wave. By Eq.(\ref{29}), this probability
is $F(\hat{i}^{'}|ii^{'})$ which coincides simply with the escaping
probability $a_n$ in the case $r=|i-\hat{i}^{'}|\rightarrow 0$.

Avalanches of radius $r$ consist of the
waves whose radius does not exceed $r$.
Therefore, the expected number of waves can be estimated as
\begin{equation} \label{50}
\langle n_i \rangle \leq c a_n^{-1}
\end{equation}
where $r^2 \sim n(\log n)^{1/3}$.
Using Eq.(\ref{35}), we get
\begin{equation} \label{51}
\langle n_i \rangle \leq c(\log n)^{1/2}
\end{equation}
or
\begin{equation} \label{52}
\langle n_i \rangle \leq c_1(\log r)^{1/2}
\end{equation}
Thus, in $d=4$, the number of waves in an avalanche of radius $r$ grows
not faster than logarithmically.

In $d=2$, the distribution of last waves is known \cite{DManna}
\begin{equation} \label{53}
P_l(R) \sim \frac{1}{R^{7/4}}
\end{equation}
The probability for the wave of radius $R$ to be the last in an avalanche
is $P_l(R)$ divided by the general distribution of waves Eq.(\ref{21}).
Then, the expected number of waves $\langle n_i \rangle$ in $d=2$
has the upper estimate
\begin{equation} \label{54}
\langle n_i \rangle \leq c r^{3/4}
\end{equation}
which is consistent with Eq.(\ref{14}) where the exact value $y=1/2$
was conjectured \cite{PKI}.

For $d > 4$, the escaping probability $a_n$ remains finite for all $n$.
Therefore, the expected number of waves in an avalanche is restricted
by a constant.

The logarithmic growth of $\langle n_i \rangle$ in $d=4$ given by
Eq.(\ref{52}) implies that the probability distributions for waves and
avalanches can differ not more than by logarithmic correction.

To estimate this correction, we take instead of the density $\rho(r)$
its upper bound Eq.(\ref{44}) if r.f.s.  Eq.(\ref{44})is less than 1,
and put $\rho(r)=1$ otherwise.
Comparing Eq.(\ref{20}) with Eq.(\ref{19})
where the new $\rho(r)$ is taken, we get the probability distribution
for waves in the form
\begin{equation} \label{55}
P(R) = \frac{\log R}{R^3}
\end{equation}
Also, we take instead of the lower bound Eq.(\ref{35}) for $a_n$  its exact
asymptotics conjectured by Lawler \cite{Law53}. Then,
\begin{equation} \label{56}
\langle n_i \rangle \sim (\log r)^{1/3}
\end{equation}
The maximal difference between avalanche and wave distributions
corresponds to a situation when all waves in an avalanche of radius $r$
have the maximal radius $r$
\begin{equation} \label{57}
\langle n_i \rangle P_{aval}(r)dr =  P(r)dr
\end{equation}
Hence, the avalanche distribution in $d=4$ can be estimated as
\begin{equation} \label{58}
P_{aval}(r) \sim \frac{(\log r)^{\gamma}}{r^3}
\end{equation}
with $2/3 \leq \gamma \leq 1$.

The probability distribution of the total number of topplings in a wave
$P(s)$ follows from Eq.(\ref{55}) provided the leading asymptotic of $s$
is $s \sim R^4/\log R$:
\begin{equation} \label{59}
P(s) \sim \frac{(\log s)^{1/2}}{s^{3/2}}
\end{equation}

In $d > 4$, the asymptotics of the avalanche and wave distributions
coincide and correspond to the mean-field behavior with the exponent
$\tau_s = 3/2$.

\section{Acknowledgements}

This work was supported by RFFR through Grant No. 99-01-00882 and
by NREL through Grant No. AAX-8-18685-01. The support and
hospitality of the Dublin Institute of Advanced Studies are gratefully
acknowledged .


\newpage
\begin{thebibliography}{}
\bibitem{BTW} P.Bak, C.Tang, and K.Wiesenfeld, Phys.Rev.Lett. {\bf 59},381
 (1987);
\bibitem{Dhar} D.Dhar, Phys.Rev.Lett. {\bf 64},1613 (1990).
\bibitem{DM} D.Dhar,S.N.Majumdar, J.Phys.A:Math.Gen. {\bf 23}, 4333 (1990).
\bibitem{JL} S.A.Janovski and C.A.Laberge, J.Phys.A:Math.Gen.
{\bf 26}, L973 (1993).
\bibitem{O} S.P. Obukhov, in {\it Random Fluctuations and Pattern Growth},
edited  by H.E.Stanley and N.Ostrovsky, Kluver, Dordrecht,
Netherlands (1988).
\bibitem{DG} A.Dias-Guilera, Europhys. Lett. {\bf 26},177 (1994).
\bibitem{Zh} Y.C.Zhang, Phys.Rev.Lett. {\bf 63}, 470 (1989).
\bibitem{CO} K.Christensen and Z.Olami,Phys.Rev. E {\bf 48},3361 (1993).
\bibitem{ZLS} S.Zapperi, K.B.Lauritsen. H.E.Stanley, Phys.Rev.Lett. {\bf 75},
4071 (1995).
\bibitem{LI}T.C.Lubensky and J.Isaacson, Phys.Rev. A {\bf 20},2130 (1979).
\bibitem{BFG}A.Bovier,J.Fr{\"o}lich,U.Glaus, Branched polymers and
dimensional reduction,in {\it Critical Phenomena,Random systems,Gauge
Theories},edited by K.Osterwalder and R.Stora, North-Holland,Amsterdam (1984).
\bibitem{TH}H.Tasaki and T.Hara, Progr.Theor.Phys.Suppl.{\bf 92},14 (1987).
\bibitem{HS}T.Hara and G.Slade, J.Stat.Phys. {\bf 59},1469 (1990).
\bibitem{MG} P.Grassberger and S.S.Manna, J.Phys.France {\bf 51},1077 (1990).
\bibitem{LU}S.L{\"u}beck and K.D.Usadel, Phys.Rev.E {\bf 56},5138 (1997).
\bibitem{L}S.L{\"u}beck, Phys.Rev.E {\bf 58},2957 (1998).
\bibitem{CMVZ}A.Chessa, E.Marinari, A.Vespignani, and S.Zapperi,
cond-mat/9802123 (1998).
\bibitem{alex}A.Vespignani, R.Dickman, M.A.Munoz, S.Zapperi,
cond-mat/9806249 v2 (1998).
\bibitem{Law}G.F.Lawler, {\it Intersection of Random Walks},Birkh{\"a}user,
Boston,Basel,Berlin (1991).
\bibitem{M} S.N.Majumdar, Phys.Rev.Lett. {\bf 68},2329 (1992).
\bibitem{Law1}G.F.Lawler, Duke Math.J.{\bf 47},655 (1980).
\bibitem{MD} S.N.Majumdar, D.Dhar, Physica {\bf A 185}, 129 (1992).
\bibitem{Harary} F.Harary, E.H.Palmer, Graphical Enumeration,
Academic Press, New York, London (1973).
\bibitem{IKP} E.V.Ivashkevich, D.V.Ktitarev and V.B.Priezzhev,
J. Phys. A:Math.Gen.  {\bf 27}, L585 (1994).
\bibitem{IKPP} E.V.Ivashkevich, D.V.Ktitarev and V.B.Priezzhev,
Physica A {\bf 209}, 347 (1994).
\bibitem{Spitzer} F.Spitzer Principles of Random Walk, New York: Van
Nostrand (1964).
\bibitem{LE}J.W.Lyklema and C.Evertsz, J.Phys.A :Math.Gen. {\bf 19},
L895 (1986).
\bibitem{Law2}G.F.Lawler, J.Phys.A :Math.Gen. {\bf 20},
4565 (1987).
\bibitem{Brod}A.Z.Broder,{\it in Proceedings of the Thirtieth
Annual IEEE Symposium on Foundations of Computer Science}
(IEEE,New York, 1989),p.442.
\bibitem{Law53}G.F.Lawler, Duke Math.J. {\bf 53},249 (1986).
\bibitem{Law86}G.F.Lawler, Commun. Math.Phys. {\bf 86},539 (1982).
\bibitem{Dup}B.Duplantier, Commun.Math.Phys. {\bf 117},279 (1988).
\bibitem{DManna}D.Dhar, S.S.Manna, Phys.Rev.E {\bf 54}, 2684 (1994).
\bibitem{PKI} V.B.Priezzhev, D.V.Ktitarev and E.V.Ivashkevich,
Phys. Rev. Lett. {\bf 76}, 2093 (1996).

\end{thebibliography}
\end{document}